\documentclass[11pt]{article}
\usepackage{mathptmx}
\usepackage{amsmath,amssymb,amsfonts,amsthm} 
\usepackage{color}
\usepackage{cite}
\usepackage{epsfig}
\usepackage{graphicx}

\usepackage{mathrsfs,latexsym}

\numberwithin{equation}{section}

\newcommand{\CC}{{\mathbb C}}
\newcommand{\RR}{{\mathbb R}}

\newcommand{\NN}{{\mathbb N}}

\newcommand{\Cc}{{\mathcal{C}}}
\newcommand{\Dc}{{\mathcal{D}}}

\newcommand{\Hc}{{\mathcal{H}}}
\newcommand{\Kc}{{\mathcal{K}}}
\newcommand{\Lc}{{\mathcal{L}}}

\newcommand{\Nc}{{\mathcal{N}}}
\newcommand{\Oc}{{\mathcal{O}}}
\newcommand{\Pc}{{\mathcal{P}}}

\newcommand{\Rc}{{\mathcal{R}}}

\newcommand{\red}[1]{{\color{black} #1}}

\newcommand{\bix}{\mbox{\boldmath $x$}}
\newcommand{\biy}{\mbox{\boldmath $y$}}

\newcommand{\fG}{{\mathfrak G}}
\newcommand{\fP}{{\mathfrak P}}
\newcommand{\fV}{{\mathfrak V}}

\newcommand{\fZ}{{\mathfrak Z}}

\def\eg{{\it e.g.\ }}
\def\ie{{\it i.e.\ }}
\def\viz{{\it viz.\ }}

\begin{document} 

%

\title{The universal C*-algebra of the electromagnetic field \\[1mm]
{\large \it To the memory of Daniel Kastler and John E. Roberts}}
\author{Detlev Buchholz${}^{(a)}$,
\ Fabio Ciolli${}^{(b)}$, \ Giuseppe Ruzzi${}^{(b)}$ \
and \ Ezio Vasselli${}^{(b)}$ \\[20pt]
\small 
${}^{(a)}$ Institut f\"ur Theoretische Physik, Universit\"at G\"ottingen, \\
\small Friedrich-Hund-Platz 1, 37077 G\"ottingen, Germany\\[5pt]
\small
${}^{(b)}$ 
Dipartimento di Matematica, Universit\'a di Roma ``Tor Vergata'' \\
\small Via della Ricerca Scientifica 1, 00133 Roma, Italy \\
}
\date{}

\maketitle

{\small 
\noindent {\bf Abstract.} A universal C*-algebra of the electromagnetic
field is constructed. It is represented in any quantum field theory  
which incorporates electromagnetism and expresses basic features of the field 
such as Maxwell's equations, Poincar\'e covariance and Einstein causality.
Moreover, topological properties of the field resulting from 
Maxwell's equations are encoded in the algebra, leading  
to commutation relations with values in its   
center. The representation theory of 
the algebra is discussed with focus on vacuum 
representations, fixing the dynamics of the field. \\[2mm]
{\bf Mathematics Subject Classification.}  \ 81V10, 81T05, 14F40  \\[2mm]
{\bf Keywords.} \ electromagnetic field, de Rham theory, local algebras,
dynamical ideals, Haag-Kastler axioms 
}

\section{Introduction}
\setcounter{equation}{0}

The general structural analysis of the electromagnetic field is a 
long-standing research topic in quantum field theory. 
Originally, this analysis was based on the 
Borchers algebra approach to quantum field theory, cf.\ 
\cite{Bongaarts}
and references quoted there. But this setting,
involving unbounded field operators, has its mathematical shortcomings
because of notorious domain problems. In particular, it does not allow
for a thorough discussion of non-regular representations, appearing for 
example in the presence of constraints \cite{ItZu},  
or of omnipresent infrared problems \cite{JaRo}. This fact 
triggered attempts to 
reformulate the theory in terms of C*-algebras \cite{Haag}. 
For the non-interacting electromagnetic field, this step can be accomplished by
proceeding to the Weyl algebra of the field, cf.\ \cite{Roepstorff} and
references quoted there. This approach works also with slight modifications 
for the electromagnetic field coupled to c-number currents  
\cite{Streater}. But the case of paramount physical interest, describing 
the coupling of the electromagnetic field to quantized relativistic 
matter, is not covered by these approaches. 

In this article, we exhibit a C*-algebra describing  
basic features of the electromagnetic field which may be taken for
granted in any relativistic quantum field theory which includes 
electromagnetism.  
Whereas the ensuing algebra does not incorporate any dynamical
law, it has a sufficiently rich structure to identify 
in its dual space states describing the vacuum. The resulting 
GNS representations fix dynamical ideals corresponding to specific
theories. The algebra thus provides a concrete C*-algebraic framework for 
the general structural analysis and physical interpretation of 
the electromagnetic field. The ideas underlying its construction 
may be known to experts; but, to the best of our knowledge, this 
approach has not yet been put on record.

To motivate the relations encoded in the algebra let us 
briefly recall the basic properties of the electromagnetic field. 
We use units where $c = \hbar = 1$ and 
consider four-dimensional Minkowski space 
$\RR^4$ with metric fixed by the Lorentz scalar product 
$x_\mu y^\mu = x_0 \, y_0 - \bix \biy$. 
The proper description of the electromagnetic field requires the 
introduction of spaces of tensor valued test functions (differential forms). 
We denote by  $\Dc_r(\RR^4)$, \mbox{$r = 0,\dots,4$},
the spaces of real test functions 
$x \mapsto f^{\mu_1\dots\mu_r}(x)$ which have compact support
in $\RR^4$ and are skew symmetric in $\mu_1\dots\mu_r$, $r \geq 2$. 
They are stable under Poincar\'e transformations
$P \doteq (y,L) \in \Pc_+^\uparrow = \RR^4 \rtimes \Lc_+^\uparrow$, given by 
$f \mapsto f_P$, \ where 
$$ f_P^{\mu_1\dots\mu_r}(x) \doteq 
L^{\mu_1}_{\ \nu_1}\cdots L^{\mu_r}_{\ \nu_r} \, f^{\nu_1\dots\nu_r}(P^{-1}x)
\, .
$$
There exist two canonical 
\red{mappings} 
between these spaces: The exterior derivative 
$d : \Dc_r(\RR^4) \rightarrow \Dc_{r+1}(\RR^4)$ is defined by 
$$ (d \, f)^{\mu_1\dots\mu_{r+1}}(x) \doteq 
- \partial^{[\mu_1}f^{\mu_2\dots\mu_{r+1}]}(x) \, ,
$$ 
where 
$\partial^\mu$ denotes the partial derivatives with respect to the 
coordinates of $x$ and the square bracket indicates anti-symmetrization.
The corresponding 
co-derivative $\delta : \Dc_r(\RR^4) \rightarrow  \Dc_{r-1}(\RR^4)$  
is given by 
$$ (\delta f)^{\mu_1  \dots  \mu_{r-1}}(x) \doteq 
- r \, \partial_{\nu} f^{\nu\mu_1\dots\mu_{r-1}}(x) \, ; 
$$
it is related to $d$ by $\delta = - \star d \star$, where  
$ \star : \Dc_r \rightarrow \Dc_{4-r}$ is the Hodge operator,  
$$
(\star f)^{\mu_1\dots\mu_{4-r}}(x) \doteq 
(1/r!) \, \epsilon_{\nu_1\dots\nu_r}{}^{\mu_1\dots\mu_{4-r}} 
f^{\nu_1\dots\nu_r}(x) \, ,
$$ 
and  $\epsilon_{\mu_1\dots\mu_4}$ the Levi-Civita tensor. 
\red{The} 
particular choice of signs in these definitions is convenient here. 

Making use of this notation, the electromagnetic field $F$ 
can be presented as an operator valued real linear map from the space 
of real test functions 
$\Dc_2(\RR^4)$ to the symmetric (hermitian) generators of some polynomial 
\mbox{*-algebra}~$\fP$, 
$$
f \mapsto F(f) = F_{\mu \nu}(f^{\mu \nu}) \, \red{, \quad f \in \Dc_2(\RR^4) \, .}
$$ 
The homogeneous Maxwell equation for the field reads
\mbox{$F(\delta h) = 0$} for $h \in \Dc_3(\RR^4)$  and
the inhomogeneous Maxwell equation 
\red{is given by} 
$j(g) = j_\mu(g^\mu) \doteq F(dg)$ for $g \in \Dc_1(\RR^4)$.
The latter 
\red{relation} 
is to be interpreted as the definition of 
the (identically conserved) current $j$ since~$F$ is regarded as being given.
Einstein causality is expressed by the condition of locality 
according to which the commutator of the electromagnetic field satisfies  
$[F(f_1), F(f_2)] = 0$
whenever the supports of $f_1, f_2 \in \Dc_2(\RR^4)$ are 
spacelike separated. 
These relations are consistent with the automorphic action of 
the Poincar\'e group on the algebra $\fP$  fixed by the mapping 
$F(f) \mapsto F(f_P)$, \ $P \in \Pc_+^\uparrow$.

It is convenient to proceed from the electromagnetic field $F$ to its 
intrinsic (gauge invariant) vector potential~$A$, given by 
$$
A(\delta f) = A_\mu((\delta f)^\mu) \doteq  F(f) \quad \mbox{for} \quad   
f \in \Dc_2(\RR^4) \, .
$$ 
Clearly, $\delta (\delta f) = 0$, $f \in \Dc_2(\RR^4)$, that is,  
$\delta f \in \Dc_1(\RR^4)$ is co-closed.  
Conversely, given any co-closed 
$g \in \Dc_1(\RR^4)$, $\delta g = 0$, 
Poincar\'e's lemma 
(cf.\ \cite[Lem.\ 17.27]{Le} and the appendix) 
implies that there exists some $f \in \Dc_2(\RR^4)$ such that $g = \delta f$,
\ie $g$ is co-exact. Moreover, the ambiguities involved in the choice 
of $f$ consist of additive terms of the 
form $\delta h$ where $h \in \Dc_3(\RR^4)$. 
Denoting by $\Cc_1(\RR^4) \subset \Dc_1(\RR^4)$ the real subspace of 
co-closed forms $g \in \Dc_1(\RR^4)$, $\delta g = 0$, 
one can therefore define for any $g  \in \Cc_1(\RR^4)$ 
\red{and corresponding co-primitive $f \in \Dc_2(\RR^4)$} the potential
$$
A(g) \doteq F(f) \, , 
\red{\quad f \in \{ f^\prime \in \Dc_2(\RR^4) : \delta f^\prime = g \}} \, .
$$   
This definition is consistent since $\delta f^\prime = g = \delta f$
implies according to Poincar\'e's lemma that  
$f^{\prime} = f + \delta h$ for some $h \in \Dc_3(\RR^4)$ 
and consequently $F(f^{\prime}) = F(f)$
by the homogeneous Maxwell equation.

In view of these facts, one can express the properties of the electromagnetic
field in terms of its intrinsic vector potential $A$. This potential 
defines a real linear map $g \mapsto A(g)$ from $\Cc_1(\RR^4)$
to the symmetric generators of the *-algebra $\fP$. 
The homogeneous Maxwell equation is satisfied by construction
and the inhomogeneous Maxwell equation now reads 
$j(g) \doteq A(\delta d g)$, $g \in \Dc_1(\RR^4)$. 
Noticing that the subspace $\Cc_1(\RR^4) \subset \Dc_1(\RR^4)$ is 
stable under the action of Poincar\'e transformations, the 
automorphic action of the Poincar\'e group on the 
algebra $\fP$ is fixed by the mappings $A(g) \mapsto A(g_P)$, 
$P \in \Pc_+^\uparrow$. 

The formulation of Einstein causality in terms of the 
intrinsic vector potential is more subtle, however, since one may not 
assume from the outset that $A$ can be extended to the space $\Dc_1(\RR^4)$ 
as a local and covariant field. These 
non-observable extensions  
require the consideration of indefinite 
metric spaces~\cite{Steinmann, Strocchi1} or of modifications of the 
*-operation~\cite{Strocchi2}, so they do not fit into 
the present setting. To avoid these 
auxiliary constructs we make use of some pertinent geometrical facts.  
Given any $g \in \Cc_1(\RR^4)$ that has support in some open 
double cone $\Oc \subset \RR^4$ it follows from 
a local version of Poincar\'e's lemma (cf.\ Appendix) that there is some 
co-primitive~$f \in \Dc_2(\RR^4)$, satisfying $\delta f = g$, that
has its support in $\Oc$ as well. The locality
of the electromagnetic field then implies  
$$ 
[A(g_1), A(g_2)] = [F(f_1), F(f_2)] = 0
$$ 
whenever \ $g_1, g_2 \in \Cc_1(\RR^4)$ 
have their supports in spacelike separated 
double cones.
If $g_1, g_2 \in \Cc_1(\RR^4)$ have their supports  
in spacelike separated but topologically non-trivial 
regions one can, however, no longer conclude that the  
commutators vanish. Yet, as is shown in the appendix, all of these  
commutators are invariant under spacetime translations, \viz  
$$
[A(g_1), A(g_2)] = [A(g_{1 \, y}), A(g_{2 \, y})]
\quad \mbox{for} \quad y \in \RR^4 \, .
$$
Hence, because of locality, they commute with all elements of the 
algebra $\fP$, \ie they are elements of its center. 
Note that the customary Gupta-Bleuler potential of the
electromagnetic field, restricted 
to the test function space $\Cc_1(\RR^4)$, provides a concrete 
representation of the algebra $\fP$ where these central elements  
vanish; yet this does not hold true within the abstract algebra.

These basic properties of the intrinsic vector potential $A$
can be recast in terms of unitary operators
that may heuristically be interpreted as its 
exponentials,  \viz 
$V(a,g) \, \widehat{=} \, \exp{(i a A(g))}$  
where $a \in \RR$, $g \in \Cc_1(\RR^4)$. As a matter of fact, this 
correspondence can rigorously 
be established in regular representations of the 
algebra $\fV$ generated by these unitaries.    
This algebra is defined in the subsequent section,
where its C*-property is also established. 
In the third section, the incorporation of 
dynamics into the framework is explained, based on 
the choice of vacuum states on the algebra~$\fV$.
This approach is illustrated by examples, clarifying its 
relation to standard field theoretic treatments. 
The article concludes with a brief summary 
and an appendix containing specific local and causal versions of  
Poincar\'e's lemma which are of relevance in the present context.

\section{The universal algebra}
\setcounter{equation}{0}

\red{The} 
construction of the universal algebra of the 
electromagnetic field 
\red{proceeds} 
from the *-algebra $\fV_0$ generated by the elements of the set
$\{ V(a,g) : a \in \RR, \, g \in \Cc_1(\RR^4) \}$
which satisfy the relations given below.  
Denoting by the symbol $\perp$ pairs of spacelike separated 
subsets of $\RR^4$ and by $\lfloor X,Y \rfloor \doteq XYX^*Y^*$ the group 
theoretic commutator of unitary operators $X,Y$, these relations read
\begin{align}
& \label{a1} V(a_1,g) V(a_2,g) = V(a_1+a_2,g) \, , 
\quad V(a,g)^* = V(-a,g) \, , \quad V(0,g) = 1 & \\
& \label{a2} 
\red{V(a_1 , \delta f_1)V(a_2 , \delta f_2) = 
V(1, a_1 \delta f_1 + a_2 \delta f_2 )} \quad 
\mbox{if} \quad \mbox{supp}\,f_1 \perp \mbox{supp}\,f_2 &  
\\
& \label{a3} 
\red{ \lfloor V(a,g), \lfloor V(a_1,g_1), V(a_2,g_2) \rfloor  
\rfloor } = 1 \quad
\mbox{if} \quad \mbox{supp}\,g_1 \perp \mbox{supp}\,g_2 \, , &
\end{align}
where $a, a_1, a_2 \in \RR$, $g, g_1 , g_2 \in \Cc_1(\RR^4)$ and
$f, f_1, f_2 \in \Dc_2(\RR^4)$. 
That is, we start with the unitary group $\fG_0$ generated by 
$\{ V(a,g) : a \in \RR, \, g \in \Cc_1(\RR^4) \}$, subject to these
relations, and proceed to the 
complex linear span of the elements of $\fG_0$ to obtain the 
*-algebra $\fV_0$.

Relation \eqref{a1}
encodes the algebraic properties of unitary one-parameter groups 
$a \mapsto V(a,g)$, expressing the idea that one deals with the exponential 
function of the smeared potential. 
Relation~\eqref{a2} combines the information that 
the electromagnetic field is linear on $\Dc_2(\RR^4)$ and local: 
since the functional equation for the exponential function 
of operators holds only if the operators commute, the additivity 
of the field manifests itself in this restricted form. 
Relation (2.3) embodies the information 
that the commutator $[A(g_1),A(g_2)]$
of the potential commutes with any other $A(g)$ whenever $g_1$, $g_2$ have
spacelike separated supports. In the special case where the supports of
$g_1, g_2$ are contained in spacelike separated double cones, it
follows from the local Poincar\'e lemma and 
relation \eqref{a2} that one has 
$$
V(a_1, g_1)V(a_2, g_2) = V(1, a_1g_1+a_2g_2) =
V(a_2, g_2)V(a_1, g_1) \, ,
$$
hence $\lfloor V(a_1, g_1), V(a_2, g_2)\rfloor = 1 $.

The algebra $\fV_0$ can be equipped with a C*-norm by a standard
construction which we recall for the convenience of the reader.
It relies on the fact that each state (\viz positive, linear 
and normalized functional) on a *-algebra
gives rise to a Hilbert space representation by the 
GNS construction, cf.\ \cite[Sec.\ III.2]{Haag}.
Since the present algebra consists of  
linear combinations of unitary operators, their respective Hilbert space 
representatives are 
bounded and their Hilbert space norm defines a  
C*-seminorm on this algebra. If the underlying state is faithful,
this seminorm is even a norm. The existence of such states 
is established in the subsequent lemma. There we 
make use of the fact that $\fV_0$ is the complex linear span of the elements of 
the unitary group $\fG_0$.

\vspace*{2mm}
\noindent \textbf{Lemma:} \ 
Let $\omega$ be the functional on the unitary group $\fG_0$
given by $\omega(V) = 0$ for 
$V \in \fG_0 \backslash \{1\}$ and $\omega(1) = 1$. 
The canonical extension of this functional to the complex linear span
of $\fG_0$  is a faithful state on $\fV_0$. 

\vspace*{2mm}
\noindent \textit{Proof:} Since the elements of $\fG_0$ form a basis of 
$\fV_0$, the linear extension of $\omega$ to $\fV_0$ is consistently 
defined by 
$\omega(c_0 1 + \sum_n c_n V_n) = c_0$,  
where $V_n \in \fG_0\backslash \{1\}$. 
Assuming \red{without loss of generality} that the unitaries $V_n$ 
are different one also obtains 
$$
\omega((c_0 1 + \sum_n c_n V_n)^* (c_0 1 + \sum_{n'} c_{n'} V_{n'})) =
|c_0|^2 + \sum_n |c_n|^2 \geq 0
$$ 
because the terms for $n \neq n'$ 
contain operators in $\fG_0\backslash \{1\}$ and therefore vanish. 
This shows that
the linear and normalized functional $\omega$ on $\fV_0$
is positive on positive elements, hence it is a
state. Moreover, since the equality sign in the above relation 
holds only for the zero element of $\fV_0$, this state is faithful,
completing the proof of the statement. \hfill $\square$ 

\vspace*{1mm}
As indicated, any state $\omega$ induces by the  
GNS construction a representation $(\pi, \Hc, \Omega)$ of $\fV_0$, 
where~$\pi$ is a homomorphism mapping $\fV_0$ into the algebra of all 
bounded operators on some Hilbert space $\Hc$ and $\Omega \in \Hc$
is a unit vector such that 
$\langle \Omega, \pi(A) \Omega \rangle = \omega(A)$, 
$A \in \fV_0$. If $\omega$ is faithful, such as the state
exhibited in the preceding lemma, the Hilbert space norm
$\| \pi(A) \|_\Hc $ of the represented operators $A \in \fV_0$ 
defines a norm on $\fV_0$ which has the C*-property.
In order not to exclude from the outset any 
representations we proceed here to the 
largest C*-norm on $\fV_0$, given by 
$$  \| A \| \doteq \sup \, \| \pi(A) \|_\Hc \, , 
\quad A \in \fV_0 \, , $$
where the supremum extends over all GNS representations of $\fV_0$. 
Note that the supremum exists since the Hilbert space 
representatives of any given 
finite sum of unitary operators are uniformly bounded.
The completion of $\fV_0$ with respect to this norm defines the 
universal C*-algebra $\fV$ of the electromagnetic field; it is  
represented in any theory incorporating electromagnetism. 

We conclude this section by showing that the algebra $\fV$ provides a 
physically significant example fitting into the general framework of 
observable algebras, established by
Haag and Kastler \cite{HaKa}. To this end,  we define
for any given open 
double cone $\Oc \subset \RR^4$ the subalgebra
$\fV(\Oc)  \subset \fV$ \red{that is} generated by the unitaries 
$\{ V(a,g) : a \in \RR, \, g \in \Cc_1(\Oc) \}$, where 
$\Cc_1(\Oc)$ denotes the subspace of co-closed forms in $\Dc_1(\RR^4)$
having support in $\Oc$. By definition, 
$\fV(\Oc_1) \subset \fV(\Oc_2)$ whenever $\Oc_1 \subset \Oc_2$,
so the assignment $\Oc \mapsto \fV(\Oc)$ defines an isotonous net of 
C*-algebras on Minkowski space $\RR^4$ 
with common identity.
Since the unitaries underlying the construction of 
$\fV$ are based on test functions with compact support,
the inductive limit of this net coincides with $\fV$. 
Moreover, as has been explained,
relation \eqref{a2} implies that the operators assigned to spacelike
separated double cones $\Oc_1, \Oc_2$ commute, in short
$[\fV(\Oc_1), \fV(\Oc_2)] = 0$. So, the net satisfies 
the condition of locality. 

In order to see that this net is also 
Poincar\'e covariant, we note that the   
relations \eqref{a1} to \eqref{a3} do not change if, for 
given $P \in \Pc_+^\uparrow$, one replaces all 
test functions by their respective Poincar\'e transforms. This 
implies that the invertible maps $\alpha$ defined on 
$\{ V(a,g) : a \in \RR, \, g \in \Cc_1(\RR^4) \}$ by 
$$ \alpha_P (V(a,g)) \doteq V(a,g_P) \, , \quad P \in \Pc_+^\uparrow \, ,$$
extend to automorphisms of the group $\fG_0$ and 
thereon to its linear span
$\fV_0$. Composing these automorphisms 
yields a representation of the Poincar\'e group,  
\red{that is} 
$\alpha_{P_1} \circ \alpha_{P_2} = \alpha_{P_1 P_2}$
and \mbox{$\alpha_P^{-1} = \alpha_{P^{-1}}$} for $P, P_1, P_2 \in \Pc_+^\uparrow$.  
Moreover, by continuity one can further extend these automorphisms 
to the \mbox{C*-algebra}~$\fV$. For the set of GNS
representations of $\fV_0$
is stable under composition with any automorphism and consequently 
$$
\| \alpha_P(A) \| = \sup \| \pi(\alpha_P(A)) \|_\Hc =
\sup \| \pi\circ \alpha_P(A) \|_\Hc = \| A \| 
\quad \mbox{for} \quad A \in \fV_0 \, .
$$ 
Finally, noticing that for any $g \in \Cc_1(\Oc)$ one has 
$g_P \in \Cc_1(P \Oc)$, it 
\red{is apparent} 
that 
$\alpha_P(\fV(\Oc)) = \fV(P \Oc)$, $P \in \Pc_+^\uparrow$,
proving the Poincar\'e covariance of the net.  

Thus the universal algebra 
$\fV$ generated by the electromagnetic field satisfies all 
Haag-Kastler axioms \cite[p. 849]{HaKa} 
with one exception: it is not a primitive algebra
since it does not have any faithful irreducible representation. In
fact, it follows from relation \eqref{a3} that $\fV$ has a 
non-trivial center. This deficiency can be resolved, however,  
by identifying suitable irreducible representations 
$(\pi, \Hc, \Omega)$ of~$\fV$. The kernel of a representation,
denoted by $\mbox{ker} \, \pi$, characterizes a two-sided ideal in $\fV$ and 
if this kernel is stable under the action of the automorphisms
$\alpha_P, P \in \Pc_+^\uparrow$, the corresponding quotient algebra 
$\fV / \mbox{ker} \, \pi$ is by construction a primitive algebra 
which satisfies all Haag-Kastler axioms. 
Even more importantly, using this device one can in principle incorporate 
any dynamics into the quotient algebra which is compatible with the basic
properties of the electromagnetic field. This issue will be discussed in 
the subsequent section. 

\section{Representations}
\setcounter{equation}{0}

All possible states of the electromagnetic field are 
described by elements 
of the dual space of the universal algebra $\fV$. We begin by 
characterizing those states and representations 
which are of primary physical interest, 
allowing it to recover from $\fV$ 
the electromagnetic field, respectively the intrinsic 
vector potential as well defined observables.

\vspace*{1mm}
\noindent \textbf{Definition:} \
Let $\omega$ be a state on $\fV$. This state is regular if all functions 
$$a_1, \dots , a_n \mapsto \omega(V(a_1,g_1) \cdots V(a_n,g_n)) \, ,
\quad  g_1, \dots ,g_n \in \Cc_1(\RR^4) \, ,$$ 
are continuous, $n \in \NN$.
It is strongly regular if all of these functions are smooth
and their derivatives at $a_1 = \dots = a_n = 0$ are bounded by
Schwartz norms of the underlying test functions (tempered).

\vspace*{2mm}
It is not difficult to see that in the GNS representation 
$(\pi, \Hc, \Omega)$ induced by a regular state $\omega$ 
the unitary one-parameter groups $a \mapsto \pi(V(a,g))$ are continuous 
in the strong operator topology. Stone's theorem therefore implies 
that there exist densely defined selfadjoint operators $A_\pi(g)$ in the
underlying Hilbert space $\Hc$ such that $\pi(V(a,g)) = e^{iaA_\pi(g)}$ for 
$a \in \RR$, $g \in \Cc_1(\RR^4)$.  So, one recovers in these 
representations the intrinsic 
vector potential. Moreover, if $\omega$ is strongly 
regular these operators have a common dense domain $\Dc \subset \Hc$, 
containing~$\Omega$, that is
stable under their action and a core for each of them. In 
particular, the correlation
functions \ $\langle \Omega, A_\pi(g_1) \cdots A_\pi(g_n) \Omega \rangle$ 
are well defined for any 
$g_1, \dots , g_n \in \Cc_1(\RR^4)$ and they are bounded by 
Schwartz norms of these test functions, $n \in \NN$.  

It follows from relation \eqref{a2} that the operators $A_\pi$ appearing
in the GNS representation induced by a strongly regular state satisfy on
their common domain $\Dc$ 
a restricted form of linearity. \red{Namely,} 
$ a_1 A_\pi(g_1) + a_2 A_\pi(g_2) = A_\pi(a_1 g_1 + a_2 g_2) $ 
whenever $g_1, g_2 \in \Cc_1(\RR^4)$ 
have their respective supports in spacelike separated 
double cones and $a_1, a_2 \in \RR$. Linearity on all of 
$\Cc_1(\RR^4)$ is ensured if the state satisfies the following
stronger  condition.

\vspace*{1mm}
\noindent \textbf{Definition:} \ 
A state $\omega$ on $\fV$ has property L if it is strongly regular and 
$$
\mbox{\large $\frac{d}{da}$} \, 
\omega(V_1 \,
V(a,g_1) V(a,g_2) V(-a,g_1 +g_2) V_2) \, \big|_{a=0} = 0 
$$
for every \ $g_1,g_2 \in \Cc_1(\RR^4)$ and \ $V_1,V_2 \in \fG_0$.

\vspace*{2mm}
We mention as an aside that in the GNS representations 
induced by states having property $L$ the operator sums \ 
$a_1 A_\pi(g_1) + a_2 A_\pi(g_2)$ \ are essentially 
selfadjoint on $\Dc$ since they 
coincide on this domain with \ $A_\pi(a_1 g_1 + a_2 g_2)$ for 
$g_1, g_2 \in \Cc_1(\RR^4)$ and $a_1, a_2 \in \RR$. 
Next, we recall the familiar characterization of 
vacuum states by their physically 
expected properties of Poincar\'e invariance and stability \cite{Haag}. 

\vspace*{1mm}
\noindent \textbf{Definition:} \
Let  $\omega$ be a pure state on $\fV$; \ $\omega$ 
is interpreted as vacuum state if for every $A,B \in \fV \,$ 
(i) \ $\omega(\alpha_P(A)) = \omega(A)$ for $P \in \Pc_+^\uparrow$,
(ii) \ $P \mapsto \omega(A \, \alpha_P(B))$ is continuous 
and (iii) \ the Fourier transforms of all functions
$x \mapsto \omega(A \, \alpha_x(B))$  on the subgroup $\RR^4 \subset 
\Pc_+^\uparrow$ of spacetime translations 
have support in the closed forward lightcone ${V}_+$
(relativistic spectrum condition). 
\vspace*{2mm}

As is well known \cite{Haag}, there exists in the GNS representation
$(\pi, \Hc, \Omega)$ induced by a vacuum state a continuous unitary
representation $U_\pi$ of the Poincar\'e group 
such that (i)~$U_\pi(P) \Omega = \Omega$, $P \in \Pc_+^\uparrow$, 
(ii)~the generators of the subgroup 
$U_\pi \upharpoonright \RR^4$ (energy-momentum) have joint spectrum in $V_+$
(\ie $\Omega$ is a ground state in all Lorentz frames) and (iii)~the 
unitaries $U_\pi$ implement the action of the Poincar\'e transformations
on observables, \viz  \ 
$U_\pi(P) \pi(A) U_\pi(P)^{-1} = \pi(\alpha_P(A))$ for any 
$P \in \Pc_+^\uparrow$ and $A \in \fV$.  
The latter relation implies that the kernel of $\pi$ is stable under 
Poincar\'e transformations. Moreover, since vacuum states are pure 
states by definition, the representation $\pi$ is irreducible. 
Hence,  proceeding to the quotient algebra $\fV / \mbox{ker} \, \pi$, 
each vacuum state on $\fV$ defines a consistent dynamical theory of 
the electromagnetic field which fulfills all Haag-Kastler axioms.
 
It follows from these remarks that the construction of theories 
involving the electromagnetic field amounts to the task of
exhibiting vacuum states in the dual space of the algebra~$\fV$. 
As a matter of fact, every vacuum state 
$\omega$ on $\fV$ is determined by its generating function
$g \mapsto \omega(V(1,g))$ on $\Cc_1(\RR^4)$. For, due to the
invariance of vacuum states under spacetime translations and the 
relativistic spectrum condition, the functions 
$x_1, \dots , x_n \mapsto 
\omega(\alpha_{x_1}(V(1,g_1)) \red{\cdots} \alpha_{x_n}(V(1,g_n)))$
extend continuously  in the variables $(x_{m + 1} - x_m)$, $m=1, \dots , n-1$, 
to the tube $(\RR^4 + i V_+)^{n-1} \subset \CC^{n-1}$
and are analytic in its interior \cite{StWi}. Moreover, for given functions
$g_m \in \Cc_1(\RR^4)$ there exist open sets of 
translations $x_m \in \RR^4$ such that the supports of the 
shifted functions $g_{m, x_m}$ are contained in spacelike separated 
double cones, $m=1, \dots , n$. Thus, the local 
Poincar\'e lemma and relation \eqref{a2} imply that 
\begin{align*}
\omega(\alpha_{x_1}(V(1,g_1)) \cdots \alpha_{x_n}(V(1,g_n)))
= \omega(V(1,g_{1,x_1}) \red{\cdots} V(1,g_{n,x_n})) \\
= \omega(V(1,g_{1,x_1} + \dots + g_{n,x_n})) \, ,
\end{align*}
where the right hand side can be determined if the generating function
is given. The left hand side can then be continued 
analytically to arbitrary 
configurations $x_m \in \RR^4$, $m=1, \dots , n$. Hence, the generating
function fixes the expectation values of vacuum states $\omega$
on the unitary group $\fG_0$, whence on all of $\fV$  by linearity 
and continuity. 

Note that if a vacuum state on $\fV$ also satisfies condition $L$ then  
the expectation values of the polynomials in the smeared 
field $F_\pi(f) = A_\pi(\delta f)$, $f \in \Dc_2(\RR^4)$, are defined in
this state 
and comply with all Wightman axioms \cite{StWi}. We make use of this fact 
in the following simple example.

\subsection{Trivial currents} 
\label{sec.noCurrent}
In this subsection, we determine all vacuum states $\omega$
on $\fV$ which lead to theories with trivial current
and have property $L$. 
The unique  result is the theory of the free 
electromagnetic field. Since this result is
obtained without any input from a classical Lagrangian, respectively
quantization scheme, it illustrates the fact that the algebra $\fV$ embodies 
fundamental physical information. 

As outlined in the introduction, the current is related to the intrinsic
vector potential by the formula $j(g) = A(\delta d g)$, 
$g \in \Dc_1(\RR^4)$. Thus in the GNS representation 
induced by a vacuum state~$\omega$ in which the current 
vanishes one has 
$$
\langle \Omega, A_\pi(\delta d g)  A_\pi(\delta d g) \Omega \rangle = 0
\, , \quad g \in \Dc_1(\RR^4) \, .
$$ 
This equality implies $ A_\pi(\delta d g) = 0$
since the vacuum vector $\Omega$ is separating 
for local operators by the Reeh-Schlieder theorem,
cf.\ \cite[Ch.\ II.5.3]{Haag}. 
Since \ $\delta d + d \delta = \square$, where
$\square$ denotes the d'Alembertian, 
one has $\delta d g = \square g$ \ for  
\mbox{$g \in \Cc_1(\RR^4) \subset \Dc_1(\RR^4)$}, so the vector potential 
fulfills the wave equation 
$A_\pi(\square g) = 0$, $g \in \Cc_1(\RR^4)$.
It then follows from locality and Poincar\'e covariance of the 
potential by standard arguments (K\"all\'en-Lehmann representation)
that its  Wightman 
two-point function coincides with that of the free field,  
\begin{align}
\label{two-point}
W(g_1,g_2) \doteq & \
\langle \Omega, A_\pi(g_1) A_\pi(g_2) \, \Omega \rangle \notag \\
= & \ c \! \int \! dp \, \theta(p_0) \delta(p^2) \, \widehat{g}_{1 \, \mu}(-p) 
\widehat{g}_2^{\, \mu}(p) \, , \quad g_1, g_2 \in \Cc_1(\RR^4) \, ,
\end{align}
where $\widehat{g}$ denotes the Fourier transform of $g$.
Rescaling the potential, one can adjust the constant in this equality 
to its conventional value $c = - (2 \pi)^{-3}$, where the 
sign is dictated by the condition of positivity of states. 
It is a remarkable consequence of this result   
that in the given vacuum representation all commutators of 
the smeared potential are multiples 
of the identity, \, $[A_\pi(g_1), A_\pi(g_2)] = c(g_1, g_2) \, 1_\Hc$,
cf.\ the arguments in~\cite{Po};
the value of $c(g_1,g_2)$ can be read off from the preceding
two-point function, $g_1, g_2 \in \Cc_1(\RR^4)$. So, one recovers  
the intrinsic vector potential of the free electromagnetic field 
in the Fock representation. The corresponding generating function
is well known, 
$$
g \mapsto \omega(V(1,g)) = 
e^{- W(g,g) /2} \, ,
\quad g \in \Cc_1(\RR^4) \, ,
$$
where $W$ is the two-point function given above. 

This special form of generating functions, depending only on a 
two-point function, is distinctive of quasifree states on $\fV$. 
Making use of the general K\"all\'en-Lehmann representation
of two-point functions, it is not 
difficult to determine all quasifree vacuum states on $\fV$
which have property $L$. Of particular interest 
is the case where the current is proportional to the 
intrinsic vector potential 
in the underlying GNS representation.
By arguments similar to those given above one finds 
that the potential then is a free massive vector field with 
two point function 
$$ W_m(g,g) = - (2 \pi)^{-3}
\int \! dp \, \theta(p_0) \delta(p^2 - m^2) \, \widehat{g}_{\mu}(-p) 
\widehat{g}^{\, \mu}(p) \, , \quad 
g \in \Cc_1(\RR^4) \, .
$$ 
The mass square   
$m^2$ is determined by the constant of proportionality between the current
and the potential. Plugging this two-point function
into the above formula yields the generating function of the 
corresponding vacuum state. 

\subsection{Classical  currents}
\label{sec.clCurrent}

Next, we discuss the cases where the electromagnetic field is coupled to 
classical currents. Such currents are simultaneously measurable  
with all other observables and are therefore described by representations 
of $\fV$ where the current operators 
$j(g) = A(\delta d g)$, $g \in \Dc_1(\RR^4)$, are affiliated with the center. 
It is clear from the outset that such representations cannot be induced
by vacuum states on~$\fV$ since the appearance of classical currents 
breaks the Poincar\'e symmetry of these states spontaneously. We are therefore 
led to consider a more general class of pure states $\omega$ on 
$\fV$ which have property $L$. The corresponding GNS representations
$(\pi,\Hc,\Omega)$ are irreducible, so their center consists of multiples 
of the identity and one has 
$\pi(V(1,\delta d g)) = e^{ij_\pi(g)} \, 1_\Hc$, $g \in \Dc_1(\RR^4)$, 
where the current $j_\pi$ is a conserved real-valued 
distribution fixed by the representation. 

\red{Whenever} the current $j_\pi$ is sufficiently regular  
it can be extended to the space 
$G_0 \, \Dc_1(\RR^4) \supset \Dc_1(\RR^4)$ obtained
by convolution of the test functions with the
time symmetric Green's function~$G_0$ of the wave equation 
(\ie half the sum of the retarded and advanced Green's 
functions). One can then define an automorphism $\gamma$ of  $\fV$,  
putting on its generating unitaries 
$$
\gamma(V(1,g)) \doteq  e^{- ij_\pi(G_0 \, g)} \, V(1,g)
\, , \quad g \in \Cc_1(\RR^4) \, ;
$$  
note that relations \eqref{a1} to~\eqref{a3} are preserved by this map. 
Composing the given representation with this automorphism
yields the representation $\pi_0 \doteq \pi \circ \gamma$ of
$\fV$ on $\Hc$. Now for $g \in \Dc_1(\RR^4)$ 
one has $\delta d g = \square \, g - d \delta g$, hence   
$j_\pi(\delta d g) = j_\pi(\square g)$ since $j_\pi$ is conserved.
As $j_\pi(G_0 \, \square g) = j_\pi(g)$ this implies
$\pi_0(V(1,\delta d g)) = 1_\Hc$, $g \in \Dc_1(\RR^4)$.
Thus the current vanishes in the representation
$\pi_0$. One may therefore consistently assume that this representation 
coincides with the vacuum representation of the free electromagnetic field. 
With this input one obtains for the
representation  $\pi = \pi_0 \circ \gamma^{-1}$ the 
generating function 
$$
\omega(V(1,g)) =  e^{ij_\pi(G_0 \, g)} \, e^{-W(g,g)/2} \, ,
\quad g \in \Cc_1(\RR^4) \, ,
$$
where $W$ is the above two-point function of the free electromagnetic
field.

\subsection{Quantum currents}

The rigorous construction of vacuum states on $\fV$ 
that describe the coupling of the electromagnetic field 
to charged quantum fields and their currents 
is a long-standing open problem.
In most approaches one proceeds from time-ordered
products of the underlying 
generating functions, denoted by $\omega(V_T(1,g))$. 
Relying on the relativistic spectrum condition,
the vacuum states $\omega$ on~$\fV$ can likewise be reconstructed 
from these functions by methods of analytic continuation. 
Heuristic candidates for the time-ordered functions 
are Feynman path integrals of the form 
$$
\omega(V_T(1,g)) \doteq Z^{-1} \! \int \! dA \, d\psi \, d\overline{\psi} \, 
e^{iS(A,\psi,\overline{\psi})} \, e^{i A(g)} \, ,
$$ 
where all charged fields $\psi, \overline{\psi}$ 
appearing in the underlying classical action $S$
are integrated out. In spite of important progress in
the rigorous construction of such integrals~\cite{AlHoMa,GlJa},   
all presently available methods of determining these 
expectations rely on renormalized perturbation 
theory, cf.\ \cite{FeHuRoWr,Scharf,FrLi} and references quoted there.
A perturbative approach to the computation of the
unordered generating functions $\omega(V(1,g))$, 
based on field equations,  
has been developed by Steinmann \cite{Steinmann}.

Since the algebra $\fV$ embodies all basic features of the 
electromagnetic field, a rigorous proof that vacuum 
states describing interaction exist in its dual space 
(possibly based on other constructive schemes) would 
be of great physical interest. On the other hand, the unlikely 
possibility that no fully consistent theory of interacting 
electromagnetic fields can be accommodated in the general framework 
of local quantum field theory would likewise manifest itself in the 
structure of the dual space of $\fV$. Thus, this algebra 
provides a solid basis for further study of this 
existence problem.

\subsection{Topological charges}

Finally, we discuss representations of the algebra $\fV$ where the 
properties of the intrinsic vector potential in non-contractible 
regions matter. The general geometrical features of nets based on 
such regions were studied in \cite{CRV1} with applications to the 
electromagnetic field and potential in \cite{CRV2}. Here, we focus on 
irreducible representations of $\fV$ where the (central)  
group theoretic commutator $\lfloor V(1,g_1), V(1,g_2) \rfloor$ 
has values different from $1$ for certain pairs 
of test functions $g_1, g_2$ with spacelike separated, linked supports.  
Since the specific topology of the supports is of importance here we 
interpret these values as topological charges.\smallskip 

At the algebraic level, let us first 
\red{note} 
that there 
\red{exist functions} 
$g_1, g_2 \in \Cc_1(\RR^4)$ with spacelike separated
supports such that \ $V(1,g_1) V(1,g_2) \neq V(1,g_2) V(1,g_1)$.
\red{For the} 
equality of such spacelike separated unitaries 
is not required in the defining relations \eqref{a1} to \eqref{a3}.
Hence if equality holds nonetheless for a given 
pair of functions, this must be a consequence of the 
condition of locality of the electromagnetic field which is encoded in 
relation \eqref{a2}. In other words, there must exist functions 
$f_1, f_2 \in \Dc_2(\RR^4)$ with $\text{supp} f_1 \perp \text{supp} f_2$
such that $g_1 = \delta f_1$ and $g_2 = \delta f_2$.
Yet, as can be inferred from the work of Roberts \cite[\S 1]{Roberts}, 
there exist pairs $g_1, g_2 \in \Cc_1(\RR^4)$, having their 
supports \red{in} spacelike separated linked 
loop-shaped regions, for which this 
condition cannot be satisfied. Hence, the corresponding 
group-theoretic commutators $\lfloor V(1,g_1), V(1,g_2) \rfloor$ 
are non-trivial unitary operators.
We denote by $\fZ \subset \fV$ the C*-algebra generated
by $\lfloor V(1,g_1), V(1,g_2) \rfloor$  for all pairs of 
functions $g_1, g_2 \in \Cc_1(\RR^4)$  with  spacelike separated
supports. According to the preceding remarks, this algebra
is non-trivial and, by relation~\eqref{a3}, contained in 
the center of $\fV$. In particular, $\fZ$ is an abelian algebra. 

Given any pure state (character) $\zeta$ on $\fZ$, this state
can be extended to some pure state $\omega_\zeta$ on $\fV$ 
by the Hahn-Banach theorem \cite[II.6.3.2]{Bl}. 
In the corresponding GNS representation 
$(\pi_\zeta, \Hc_\zeta , \Omega_\zeta)$ 
all elements of the algebra $\fZ$ are represented by 
multiples of the identity (the topological charges) 
whose values depend on the choice of the state $\zeta$.
It is an open problem whether for some suitable non-trivial choice of 
$\zeta$ there exist pure extensions $\omega_\zeta$ which are 
Poincar\'e invariant. However, 
states with non-trivial topological charges are in general not regular.

\section{Summary} 

In the present investigation, we have constructed a universal C*-algebra
$\fV$ of the electromagnetic field whose basic features  
are encoded in the defining relations \eqref{a1} to \eqref{a3}. Even though 
this algebra does not contain any dynamical information, it has a 
sufficiently rich structure to identify in its dual space all 
possible vacuum states of the field which depend on its particular   
coupling to charged matter. The GNS representations 
$\pi$ resulting from these states allow to construct selfadjoint 
generators of the spacetime transformations,
comprising the desired dynamical information, and the corresponding quotient 
algebras $\fV / \mbox{ker} \, \pi$ of observables satisfy all 
Haag-Kastler axioms. In the simple cases of trivial or classical currents,  
one can determine the underlying states  
without relying on any further input, such as a classical Lagrangian 
or canonical commutation relations. 
This fact shows that the universal algebra subsumes essential physical 
information. The still pending rigorous construction of interacting 
theories of the electromagnetic field amounts to the identification  
of appropriate vacuum states in the dual space of  $\fV$. It 
therefore calls for further study of this algebra. 

Apart from these constructive aspects, the universal algebra provides 
a solid basis for the structural analysis, the physical interpretation
and the classification of theories incorporating electromagnetism.
For example, based on the Haag-Kastler axioms, a general scattering theory for 
the electromagnetic field has been developed in \cite{Bu1}, the 
notorious infrared problems appearing in this context were analyzed 
in \cite{Bu2,FrMoSt} and, more recently, the possible charge structures  
and statistics appearing in theories of the electromagnetic field were 
determined in \cite{BuRo}. 
Thus, the universal algebra $\fV$ is 
a concrete and physically significant example fitting into the 
general algebraic framework of relativistic quantum field theory. 

Furthermore, the universal algebra $\fV$ seems to fully encode the geometrical 
features underlying gauge theories. 
In particular, the locality properties 
of the electromagnetic field, encoded in 
the commutation relations \eqref{a2} and \eqref{a3} of the 
intrinsic vector potential, lead to 
the emergence of new types of topological charges that can be described by 
cohomological invariants associated with linked commutators.  
We believe that these aspects are not confined to the 
electromagnetic field but should be present also in non-Abelian gauge 
theories. Thus, further investigations of these structures seem 
warranted. 

\section*{Acknowledgment}
DB  gratefully acknowledges the hospitality and financial support 
extended to him by the University of Rome ``Tor Vergata'' which made
this collaboration possible. 
FC and GR are partially supported by PRIN-2010/2011.
FC is also supported by the ERC Advanced Grant 227458 OACFT 
``Operator Algebras and Conformal Field Theory''.

\section*{Appendix}
We show in this appendix that the commutator of 
the intrinsic vector potential, smeared with test functions that have
spacelike separated supports, lies in the center of the polynomial 
algebra~$\fP$. The argument is based on refinements of Poincar\'e's lemma 
that put emphasis on the support properties of the co-primitives. 
For the convenience of the reader, we outline the proofs of 
these basic results, noting that the subsequent facts 
adopted from differential geometry and algebraic 
topology on $\RR^4$ carry over to Minkowski
space since they do not depend on the choice of a metric. 

\vspace*{2mm} 
\noindent \textbf{Local Poincar\'e Lemma:}  \
Let $g \in \Cc_1(\RR^4)$, where $\, \mbox{supp} \, g$ is contained in 
a given open star-shaped region $\Rc \subset \RR^4$ (\eg a double cone). 
There exists $f \in \Dc_2(\RR^4)$ with 
$\, \mbox{supp} \, f \subset \Rc$ such that $\delta f = g$.

\vspace*{2mm} 
\noindent \textit{Proof:} \
Making use of the fact that \ 
$\star \star \upharpoonright \Dc_r(\RR^4) 
= (-1)^{r+1} \upharpoonright \Dc_r(\RR^4)$, $r = 0, \dots , 4$, 
one obtains
$d \star f = - \star \, \delta f$ for $f \in \Dc_2(\RR^4)$.
It is then apparent that finding a co-primitive for $g$ is equivalent to 
finding a primitive $f'$ for \ $h \doteq - \star g \in \Dc_3(\RR^4)$, 
\ie $d f' = h$. Note that $h$ has the same support as $g$ since the 
Hodge operator does not change supports.
Since $d h = \star \, \delta g = 0$ it follows from the
compact Poincar\'e lemma \cite[Lem.\ 17.27]{Le} that there is 
some $f'' \in \Dc_2(\RR^4)$ such that $df'' = h$. 
To modify $f''$ so as to obtain a co-primitive supported in 
$\Rc$ we make use of the fact that 
$\mbox{supp} \, h \subset \Rc$ is compact. Hence, there is an open 
neighborhood $\Nc$ of $\RR^4 \backslash \Rc$ 
such that $df'' = h = 0$ on~$\Nc$.
For convenience, we choose $\Nc  \sim \RR^4 \backslash \Rc$
such that $\mbox{supp} \, g \subset \RR^4 \backslash \Nc$, 
where~$\sim$ denotes homotopy equivalence \cite[p.\ 614]{Le}.   
Since $\Rc$ is star-shaped it is homotopic to a point $o \in \Rc$ and 
one has $\Nc \sim \RR^4 \backslash \Rc \sim \RR^4 \backslash \{ o \}$. 

Now the corresponding de Rham cohomology groups 
of homotopic manifolds are isomorphic, cf. 
\cite[Thm.\ 17.11]{Le}, so it follows
from \cite[Cor.\ 17.23]{Le} that the second de Rham cohomology 
group of $\Nc$ is trivial. In other words, every 
closed two-form $f''$ on  $\Nc$ is exact and  
there is some smooth one-form $g''$ such that $d g'' = f''$
on  $\Nc$. Picking some smooth characteristic function 
$\chi$ with \ $\chi \upharpoonright \RR^4 \backslash \Rc = 1$
and \ $\chi \upharpoonright  \RR^4 \backslash \Nc = 0$ we put \  
$f' = f'' - d \, \chi g''$. Clearly, 
$\mbox{supp} \, f' \subset \Rc$ and $df' = df'' = h$.
Thus, $f \doteq \! - \star \! f'$ is the desired co-primitive of $g$.
\hfill $\square$

\vspace*{2mm}
\noindent \textbf{Causal Poincar\'e Lemma:} \
Let $g \in \Cc_1(\RR^4)$ and $\Oc$ an open double cone 
\red{satisfying} 
$\Oc \perp \mbox{supp} \, g$.
There exists $f \in \Dc_2(\RR^4) $ such that $\delta f = g$ and 
$ \mbox{supp} \, f \perp \Oc$.

\vspace*{2mm}
\noindent \textit{Proof:}
Since the support of $g$ is compact there are open double cones
$\Oc_2 \supset \Oc_1 \supset \overline{\Oc}$ such that 
$\mbox{supp} \, g \subset {\Oc_1^{\, \prime} \bigcap \Oc_2}$; we put
$\Kc \doteq  \overline{\Oc_1^{\, \prime} \bigcap \Oc_2}$,
where the bar denotes closure, cf.\ the figure. Note that the 
\begin{figure}[h] \label{figure}
\hspace*{42mm}
\epsfig{file=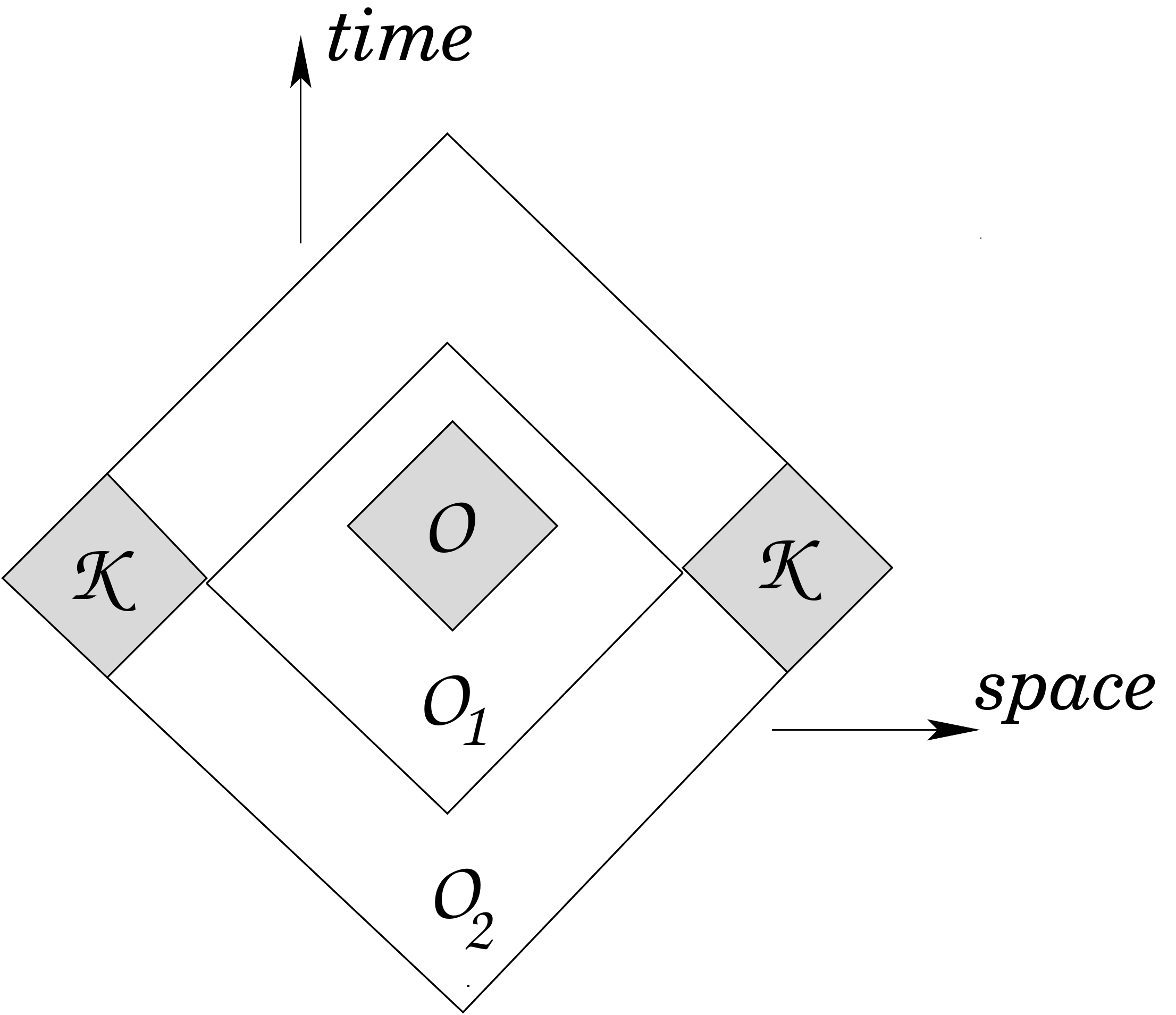,height=50mm}

\centerline{Fig.\ Collar-shaped region $\Kc$ in the spacelike complement 
of $\Oc$}
\end{figure}
collar-shaped region $\Kc$ is simply connected in four spacetime dimensions.
Since $\mbox{supp} \, g \subset \Oc_2$ and 
$\Oc_2$ is star-shaped, there exists 
according to the preceding lemma a co-primitive 
$f' \in \Dc_2(\RR^4)$, \ie $\delta f' = g$, which has
support in~$\Oc_2$. 

To exhibit a co-primitive $f$ of $g$ 
which has its support in a neighborhood of $\Kc$ we have to rely on methods
of algebraic topology \cite{Br,Re}. \red{To this end we} consider the function  
$f'' \doteq - \star  f' \in \Dc_2(\RR^4)$ and note that 
$d f'' = \star \delta f' = \star g = 0$ on 
$\RR^4 \backslash \Kc$. Since $\Kc$ is compact it follows 
\red{from} 
the  Alexander duality theorem 
\cite[Cor.\ 8.6]{Br} that the second (real) homology group
of $\RR^4 \backslash \Kc$ is isomorphic to the first cohomology
group of $\Kc$, 
$H_2(\RR^4 \backslash \Kc) \thickapprox H^1(\Kc)$; note that these
groups coincide with the corresponding reduced groups in the case at hand.
Moreover, since $\Kc$ is simply connected we have 
$0 = H^1(\Kc) \thickapprox H_2(\RR^4 \backslash \Kc) \thickapprox
H^2(\RR^4 \backslash \Kc)$,  where the latter isomorphism 
relies on the fact that the coefficient group of interest here is $\RR$,
cf.\ \cite[Thm.\ 6.9]{Re}. 
Finally, by de~Rham's Theorem we obtain 
$H^2_{dR}(\RR^4 \backslash \Kc) \thickapprox H^2(\RR^4 \backslash \Kc) = 0$,
cf.\ \cite[Thm.\ 6.9]{Br}, 
hence every closed two-form on $\RR^4 \backslash \Kc$ is exact.

So we conclude that there exists a smooth one-form $g''$ 
such that $d g'' = f''$ on $\RR^4 \backslash {\Kc}$. 
Choosing some open bounded region $\Nc$,  
${\Kc} \subset 
\Nc \subset \Oc'$, and a smooth characteristic function $\chi$
which satisfies $\chi \upharpoonright \Kc = 1$ and
$\chi \upharpoonright \RR^4 \backslash \Nc = 0$,
we define $f''' \doteq f'' - d (1 - \chi) g''$. Then,   
$\, d f''' = d f'' = \star g$ on $\RR^4$ and 
$\mbox{supp} \, f''' \subset \Nc$.  
Hence, $f \doteq \star f'''$ is a co-primitive of
$g$ which has its support in $\Nc \subset \Oc'$, completing the 
proof. \hfill $\square$

After these preparations, we turn now to the analysis of 
commutators of the intrinsic vector potential, smeared with 
test functions having arbitrary spacelike separated supports. 
Let $g \in \Cc_1(\RR^4)$ and let $g_y$ be its 
translate for given $y \in \RR^4$. As one sees by a straightforward computation,
there exists a co-primitive $f_y \in \Dc_2(\RR^4)$ of the 
difference $(g_y - g) \in \Cc_1(\RR^4)$, given by 
$$ x \mapsto f_y^{\mu_1 \mu_2}(x)   
\doteq  (1/2) \int_0^1 \! dt \, (y^{\mu_1} g^{\mu_2}(x - ty) 
- y^{\mu_2} g^{\mu_1}(x - ty)) \, . 
$$
It has support in the cylindrical region 
$\{ \mbox{supp} \, g + t y : 0 \leq t \leq 1 \}$ which,
for sufficiently small $y$, is contained in an arbitrarily small 
neighborhood of the support of $g$. 
Now let \mbox{$g_1, g_2 \in \Cc_1(\RR^4)$}
have spacelike separated supports. Then, for sufficiently small
translations $y \in \RR^4$, there is a co-primitive 
$f_{2 y} \in \Dc_2(\RR^4)$ of
$(g_{2 y} - g)$ with $\mbox{supp} \, f_{2 y} \perp \mbox{supp} \, g_1$.
By a partition of unity one can decompose this
co-primitive into a sum $f_{2 y} = \sum_{m=1}^n f_{2 y,m}$
of elements $f_{2 y,m} \in \Dc_2(\RR^4)$ which 
have their supports in double cones 
$\Oc_m \perp \mbox{supp} \, g_1$, $m = 1, \dots , n$. 
Thus,
\begin{align*}  
[A(g_1), A(g_{2 y} -g_2)] = & \ [A(g_1), A(\delta f_{2 y})] \\
= & \sum_{m=1}^n \, [A(g_1), A(\delta f_{2 y, m})]  
=    \sum_{m=1}^n \, [A(\delta f_{1,m}), A(\delta f_{2 y, m})] \, ,
\end{align*}
where in the last equality the causal Poincar\'e lemma was used 
according to which 
there exist co-primitives $f_{1,m} \in \Dc_2(\RR^4)$ of $g_1$ 
such that $\mbox{supp} \, f_{1,m} \perp \Oc_m$,
$m = 1, \dots , n$.
Because of locality of the electromagnetic field \red{one has}
$$[A(\delta f_{1,m}), A(\delta f_{2 y, m})] = 0
\, , \quad m = 1, \dots , n \, ,
$$
and consequently $ [A(g_1), A(g_{2 y})] = [A(g_1), A(g_2)]$
for small $y$. Applying the same argument to the
translates of $g_1$ one finds that 
$ [A(g_{1 y}), A(g_{2 y})] = [A(g_1), A(g_2)] $ for sufficiently small
$y$. This equality extends to arbitrary translations 
$y \in \RR^4$ by iteration. 
It then follows from locality and the Jacobi identity that the commutator
$[A(g_1), A(g_2)]$ commutes with any other operator $A(g) \in \fP$
\ if \ $g_1, g_2 \in \Cc_1(\RR^4)$ and \
$\mbox{supp} \, g_1 \perp \mbox{supp} \, g_2$. 
Hence, it lies in the center of $\fP$, completing the proof.

\end{document}